\documentclass[12pt]{article}
\usepackage{a4wide}

\usepackage[dvipdfmx,hiresbb]{graphicx} 
\usepackage{mediabb}                     

\usepackage{color}
\definecolor{refkey}{rgb}{1,0,1}
\definecolor{labelkey}{rgb}{1,0,1}

\newcommand{\magenta}[1]{{\color{black}#1}}

\newcommand{\al}[1]{\begin{align}#1\end{align}}

\newcommand{\paren}[1]{\left(#1\right)}

\newcommand{\sqbr}[1]{\left[#1\right]}

\usepackage{amsmath,amssymb}
\newcommand{\df}{\text{d}}

\usepackage{epsf}

\newcommand{\GeV}{\ensuremath{\,\text{GeV} }}

\newcommand{\eV}{\ensuremath{\,\text{eV} }}

\newcommand{\nn}{\nonumber\\}

\begin{document}

\title{\vbox{
\baselineskip 14pt
\hfill \hbox{\normalsize KUNS-2493, OU-HET/812}
} \vskip 1cm
\bf \large Predictions on mass of Higgs portal scalar dark matter\\
from Higgs inflation and flat potential\vskip 0.5cm
}
\author{
Yuta~Hamada,\thanks{E-mail: \tt hamada@gauge.scphys.kyoto-u.ac.jp}~
Hikaru~Kawai,\thanks{E-mail: \tt hkawai@gauge.scphys.kyoto-u.ac.jp}~ 
and Kin-ya~Oda\thanks{E-mail: \tt odakin@phys.sci.osaka-u.ac.jp}\bigskip\\
{\it \normalsize
${}^{*\dagger}$Department of Physics, Kyoto University, Kyoto 606-8502, Japan}\smallskip\\
\it \normalsize
${}^\ddag$Department of Physics, Osaka University, Osaka 560-0043, Japan
}
\date{\today}


\maketitle

\abstract{\noindent \normalsize
We consider the Higgs portal $Z_2$ scalar model as the minimal extension of the Standard Model (SM) to incorporate the dark matter.
We analyze this model by using the two-loop renormalization group equations.
We find that the dark matter mass is bounded to be lighter than 1000\,GeV within the framework that we have proposed earlier, where the Higgs inflation occurs above the SM cutoff $\Lambda$, thanks to the fact that the Higgs potential becomes much smaller than its typical value in the SM: $V\ll\Lambda^4$.
We can further fix the dark matter mass to be $400\GeV< m_\text{DM}<470\GeV$
if we impose that the cutoff is at the string scale $\Lambda\sim10^{17}\GeV$ and that the Higgs potential becomes flat around $\Lambda$, as is required by the multiple point principle or by the Higgs inflation at the critical point.
This prediction is testable by the dark matter detection experiments in the near future.
In this framework, the dark matter and top quark masses are strongly correlated, which is also testable.
}



\normalsize
\newpage

\section{Introduction and summary}
Recent discovery of the Higgs boson~\cite{Aad:2012tfa,Chatrchyan:2012ufa} determines all the parameters 
of the Standard Model (SM) except for the neutrino sector. Since we have seen nothing beyond the SM at the Large Hadron Collider so far, it becomes more important to consider a scenario where SM is not much altered up to a very high scale such as the string (Planck) scale, around $10^{17}$ ($10^{18}$) GeV.

If we extrapolate the obtained SM parameters toward this scale assuming no new physics, we find a curious situation that both the Higgs self coupling $\lambda$ and its beta function $\beta_\lambda$ become very small~\cite{Holthausen:2011aa,Bezrukov:2012sa,Degrassi:2012ry,Alekhin:2012py,Masina:2012tz,Hamada:2012bp,Jegerlehner:2013cta,Buttazzo:2013uya}, as well as the bare Higgs mass $m_B^2$~\cite{Hamada:2012bp,Jegerlehner:2013cta,Jegerlehner:2013nna,Hamada:2013cta,Bian:2013xra,Jones:2013aua}; see also Ref.~\cite{Alsarhi:1991ji} for the earlier two-loop computation of $m_B^2$.

This fact may give us a hint for a deeper understanding of the Planck scale physics, and is very interesting.
So far, there are several proposals of the physics behind this fact: the multiple point principle (MPP)~\cite{Froggatt:1995rt,Froggatt:2001pa,Nielsen:2012pu}, the asymptotic safety~\cite{Shaposhnikov:2009pv}, the scale invariance~\cite{Meissner:2007xv,Aoki:2012xs,Khoze:2014xha,Kobakhidze:2014afa}, the maximum entropy principle~\cite{Kawai:2011qb,Kawai:2013wwa,Hamada:2014ofa,Kawana:2014vra,Hamada:2014xra}, the hidden duality and symmetry~\cite{Kawamura:2013kua,Kawamura:2013xwa}, etc.
There are also bottom-up approaches to extend the SM to realize such a structure at high scales~\cite{Meissner:2006zh,Foot:2007iy,Iso:2009ss,Iso:2009nw,Iso:2012jn,Haba:2013lga,Hashimoto:2014ela,Chankowski:2014fva,Kawana:2014zxa,Kawana:2015tka}.
The flat potential, namely the vanishing quartic coupling around the string scale, is interesting not only theoretically but also phenomenologically.

Recently, the authors have pointed out the possibility that this flat potential can be used for the cosmic inflation~\cite{Hamada:2013mya}.
More concrete model can be found in Refs.~\cite{Hamada:2014iga,Bezrukov:2014bra}, where the inflation scale is sufficiently large to explain the recently observed B-mode polarization, i.e.\ the large tensor-to-scalar ratio, by the BICEP2 experiment~\cite{Ade:2014xna}.
See also Refs.~\cite{Kamada:2013bia,Jegerlehner:2014mua,Burgess:2014lza,Lee:2014spa,Gong:2014cqa,Okada:2014lxa,Fairbairn:2014nxa,DiBari:2014oja,Oda:2014rpa,Cheng:2014bta,Enqvist:2014tta,Enqvist:2014bua,Feng:2014naa,Bamba:2014mua,Ren:2014sya,Bartrum:2013fia,Bastero-Gil:2014jsa,Bastero-Gil:2014oga,Hosotani:1985at,Chakravarty:2013eqa}.

However, it is certain that we need to extend the SM since it does not contain a dark matter (DM).
It is probable that
Nature chooses minimal extension among many candidate models since it has not shown us any symptom of new physics at the Large Hadron Collider so far.
One of the most minimal models is the gauge singlet scalar model with the $Z_2$ symmetry~\cite{Silveira:1985rk,McDonald:1993ex,Burgess:2000yq,Davoudiasl:2004be,Patt:2006fw,Grzadkowski:2009mj,Drozd:2011aa,Haba:2013lga}.
This model is well studied so that the relation between the mass of the DM and its coupling to the Higgs is constrained to yield the correct thermal abundance of the DM; the bounds from the collider and (in)direct detection experiments are also examined; see Ref.~\cite{Cline:2013gha} for the latest analysis.


In this paper, we study the modification of the running of the SM parameters in this model.
Particularly, we examine the running of the Higgs quartic coupling to check its consistency with the Higgs inflation above the SM cutoff~\cite{Hamada:2013mya}.\footnote{
\magenta{One may think of the possibility of using this singlet dark matter as the inflaton. However, its quartic coupling does not become small in the minimal model since it does not have a Yukawa coupling. Therefore we do not consider the possibility of using the scalar dark matter as the inflaton. It may be interesting to pursue such a possibility in the model where the Higgs portal scalar has a Yukawa coupling to extra fermion; see e.g.\ Ref.~\cite{Ko:2014eia} and references therein.}
}

We show that the DM mass must be smaller than 1000\,GeV in order for the SM Higgs potential to be smaller than the inflation energy $10^{65}\,\text{GeV}^4$ all the way up to the string scale $\sim10^{17}$\,GeV.
It is interesting that the allowed region is testable in future direct detection experiments~\cite{Cline:2013gha}.

We further get the DM mass $400\GeV< m_\text{DM}<470\GeV$
if we impose the flatness of the Higgs potential $\lambda\simeq\beta_\lambda\simeq0$ around the string scale $10^{17}\GeV$, as is expected from the MPP or is required in the Higgs inflation at the critical point~\cite{Hamada:2013mya,Bezrukov:2014bra}.

This paper is organized as follows:
In Section~\ref{model}, we review the $Z_2$ scalar model and its allowed region of the parameter space.
In Section~\ref{constraint}, we briefly explain the flat potential Higgs inflation, and its constraint on the parameter space. We will see that the DM mass is strongly constrained.
In Section~\ref{summary}, we conclude this paper.

\section{$Z_2$ scalar model}\label{model}
We add a gauge singlet real scalar $S$ to the SM.
We further impose the $Z_2$ symmetry under which the SM fields are even and $S$ is odd.
This $Z_2$ assignment prohibits the decay of $S$ into the SM particles, making it stable.
The Lagrangian is:
\al{\label{Lagrangian}
\mathcal{L}=\mathcal{L}_{\text{SM}}
+\frac{1}{2}(\partial_{\mu}S)^2-\frac{1}{2}m_S^2S^2
-\frac{\rho}{4!}S^4-\frac{\kappa}{2}S^2 H^\dagger H.
}

Let us first see the behavior of $S$ as the DM.
The mass eigenvalue is
\al{
m_\text{DM}^2
	&=	m_S^2+{\kappa v^2\over2},
}
where $v\simeq246$\,GeV is the Higgs vacuum expectation value (VEV).
The singlet $S$ takes part in the thermal bath of the SM sector through the coupling $\kappa$ to the Higgs.
After this interaction is frozen out, the abundance of $S$ is fixed.
Therefore, the abundance solely depends on the DM mass $m_\text{DM}$ and the coupling $\kappa$, and is independent of the self coupling $\rho$.

For $m_\text{DM}\gtrsim m_h$ ($\simeq126$\,GeV), the condition for the correct thermal abundance is~\cite{Cline:2013gha}
\al{\label{thermal}
\log_{10}\kappa
	&\simeq	-3.63+1.04\log_{10}{m_\text{DM}\over\text{GeV}},
}
which is roughly $m_\text{DM}\sim 330\GeV\times\frac{\kappa}{0.1}$.
On the other hand, the light DM is constrained from the invisible decay width of the Higgs at the 95\% C.L.~\cite{Cline:2013gha}:
\al{
m_\text{DM}> 53\GeV.
}
The direct detection bound from XENON100 (2012) leaves two allowed regions within 90\% C.L.~\cite{Cline:2013gha}:
\al{
m_\text{DM}	&< 65\GeV,	&
m_\text{DM}
	&>
		80\GeV.
}

Next we discuss how the existence of $S$ affects the renormalization group (RG) running of the SM parameters.
For simplicity, we employ the one-loop beta functions.
Since $S$ is gauge singlet and does not have a Yukawa coupling to the SM fermions at the one-loop level, it does not affect the beta function of the gauge and Yukawa couplings, whereas modifying the running of the Higgs quartic coupling $\lambda$ via the coupling $\kappa S^2H^\dagger H$. As the result, we get
\al{
\beta_\lambda=\beta_{\lambda,\text{SM}}+\frac{1}{16\pi^2}\frac{1}{2}\kappa^2,
}
where $\beta_\lambda=\df\lambda/\df\ln\mu$.
Similarly, we obtain the modification to the bare Higgs mass:
\al{
m_B^2=m_{B,\text{SM}}^2-\frac{1}{2}\kappa I_1,
}
where
\al{
\quad I_1
	&=	\int^\Lambda \frac{\df^4p}{(2\pi)^4} \frac{1}{p^2}
	=	{\Lambda^2\over16\pi^2},
}
with the integral being over the Euclidean four momenta.

The running of the new parameters $\kappa$ and $\rho$ are
\al{
\frac{\df\kappa}{\df t}	
        &=\frac{\kappa}{16\pi^2}\left(12\lambda+ \rho+4\kappa+6y_t^2-\frac{3}{2}g_Y^2 -\frac{9}{2}g_2^2\right),\nn
\frac{\df\rho}{\df t}	
        &=\frac{1}{16\pi^2}\left(3\rho^2+12 \kappa^2\right),
}
and the bare mass of $S$ becomes
\al{
m_{S,B}^2&=-(2\kappa+\rho)I_1.
}

In this paper, we consider the case $\kappa>0$ since we do not want to have a VEV for $S$.
For a mildly small value of $\kappa$ and $\rho$, the top Yukawa contribution $y_t^2$ dominates in the running of $\kappa$, and hence $\kappa$ monotonically increases with the renormalization scale.
Since $S$ has only quartic couplings of the form $\rho S^4$ and $\frac{\kappa}{2} S^2 |H|^2$, the coupling $\rho$ also increases monotonically.

Here,  we use the two-loop RGEs summarized in Appendix~\ref{RGE}.
The initial condition for the SM parameters follow from Ref.~\cite{Buttazzo:2013uya}:\footnote{
The latest combined result for the top quark mass is $173.34\pm0.76\GeV$~\cite{ATLAS:2014wva}.
In this paper, we use the capital letter for the top quark mass to indicate that it is the pole mass.
Note that there can be a discrepancy between the pole mass $M_t$ and the one measured at the hadron colliders~\cite{ATLAS:2014wva}; see e.g.\ Refs.~\cite{Alekhin:2012py,Hamada:2013cta}.
The latter is obtained as an invariant mass of the color singlet final states, whereas the former is a pole of a colored quark. At the hadron colliders, the observed $t\bar t$ pair is dominantly color octet, and there may be discrepancy of order 1--2\,GeV caused by drawing extra lines in the 
Feynman diagram to make the singlet final states. We thank Yukinari Sumino on this point. See also Ref.~\cite{Horiguchi:2013wra}.
(This footnote is shared with the version 2 of Ref.~\cite{Hamada:2014iga} to appear.)
}
\al{
g_2(M_t)
	&=	0.64822+0.00004\left(\frac{M_t}{\GeV}-173.10\right)+0.00011\left(\frac{m_W-80.384\GeV}{0.014\GeV}\right),\\
g_Y(M_t)
	&=	0.35761++0.00011\left(\frac{M_t}{\GeV}-173.10\right)+0.00021\left(\frac{m_W-80.384\GeV}{0.014\GeV}\right),\\
g_3(M_t)
	&=	1.1666+0.00314\left(\frac{\alpha_S(m_Z)-0.1184}{0.0007}\right)-0.00046\left(\frac{M_t}{\GeV}-173.10\right),\\
y_t(M_t)
	&=	0.93558+0.00550\left(\frac{M_t}{\GeV}-173.10\right)
-0.00042\left(\frac{\alpha_S(m_Z)-0.1184}{0.0007}\right)\nn
	&\quad
	-0.00042\left(\frac{m_W-80.384\GeV}{0.014\GeV}\right)
\pm0.00050_\text{th},\\
\lambda(M_t)
	&=	0.12711+0.00206\left(\frac{m_h}{\GeV}-125.66\right)-0.00004\left(\frac{M_t}{\GeV}-173.10\right)
\pm0.00030_\text{th}.
}
On the other hand, we can freely choose the parameters for $S$.
Hereafter, we put subscript 0 on the parameters at the top mass scale $\mu=M_t$, and we treat $\kappa_0$ and $\rho_0$ as free input parameters. Though the DM mass is different from the top mass, we neglect this small threshold correction and use the RGEs shown in Appendix~\ref{RGE} from $\mu=M_t$.

\begin{figure}[tn]
\begin{center}
\hfill
\includegraphics[width=.4\textwidth]{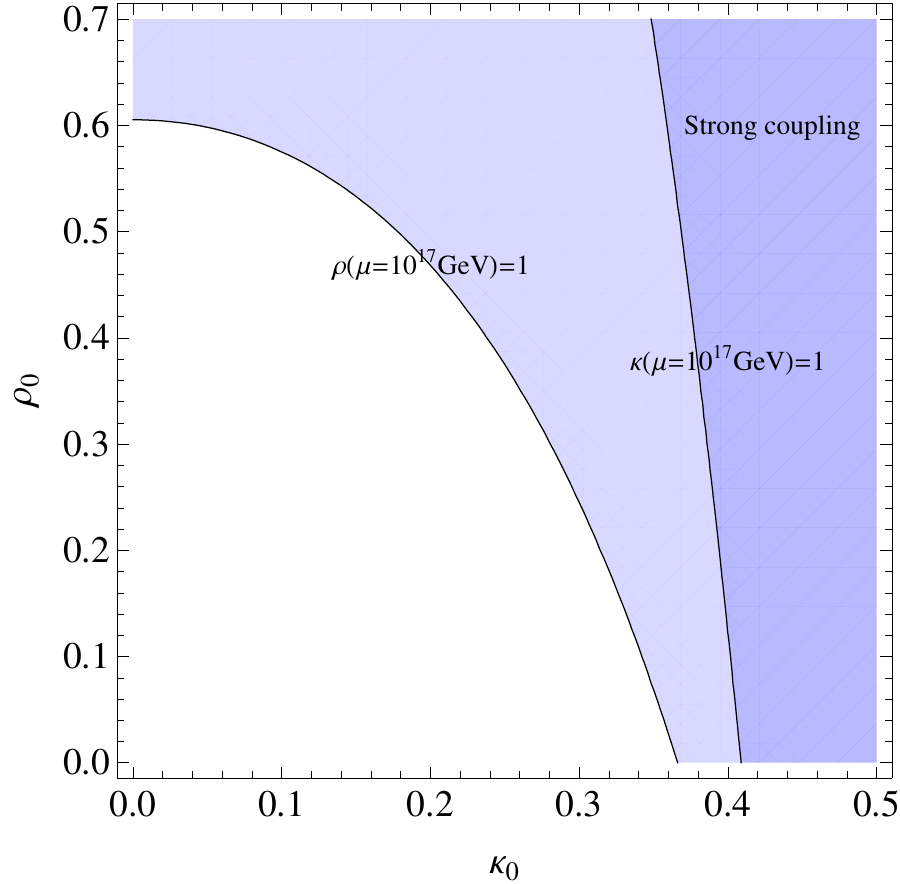}
\hfill\mbox{}
\caption{
The black solid lines represent
$\kappa=1$ and $\rho=1$ at $\mu=10^{17}$\,GeV, for $M_t=173\GeV$, $\alpha_s=0.1184$, and $m_h=126\GeV$.
Right of the upper (lower) line is the region where $\kappa$ ($\rho$) becomes large at $10^{17}\GeV$.
These solid lines hardly move when we vary the SM parameters within the experimental errors. 
}\label{perturbativity}
\end{center}
\end{figure}

\begin{figure}[tn]
\begin{center}
\hfill
\includegraphics[width=.49\textwidth]{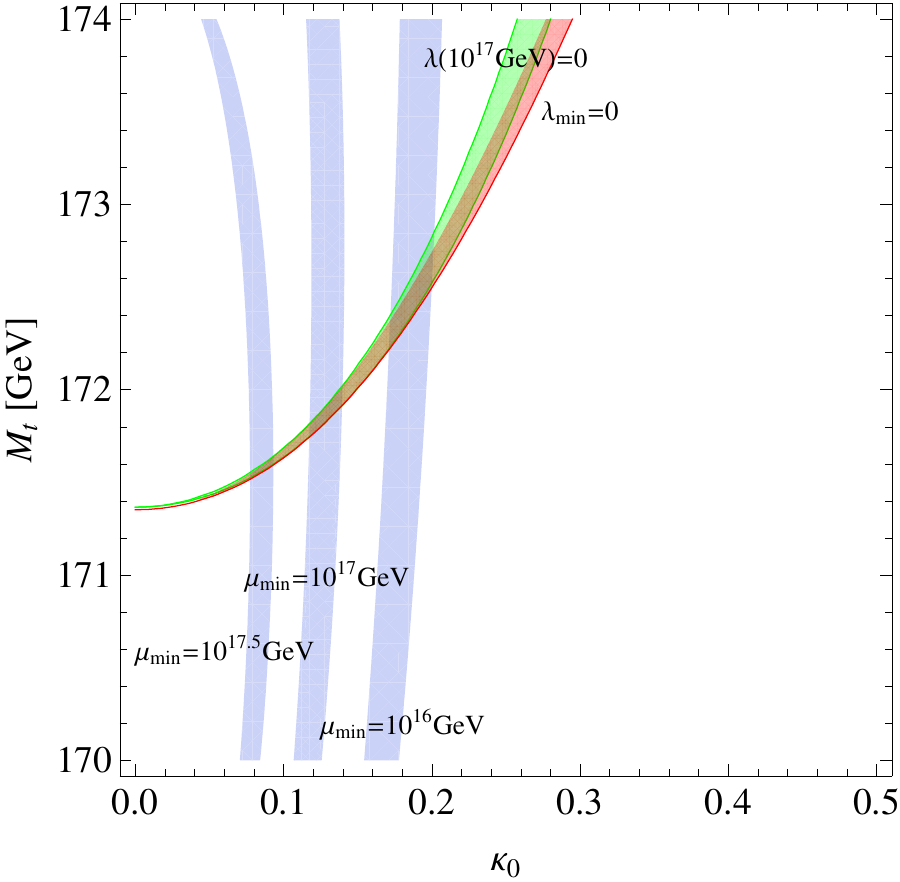}
\includegraphics[width=.49\textwidth]{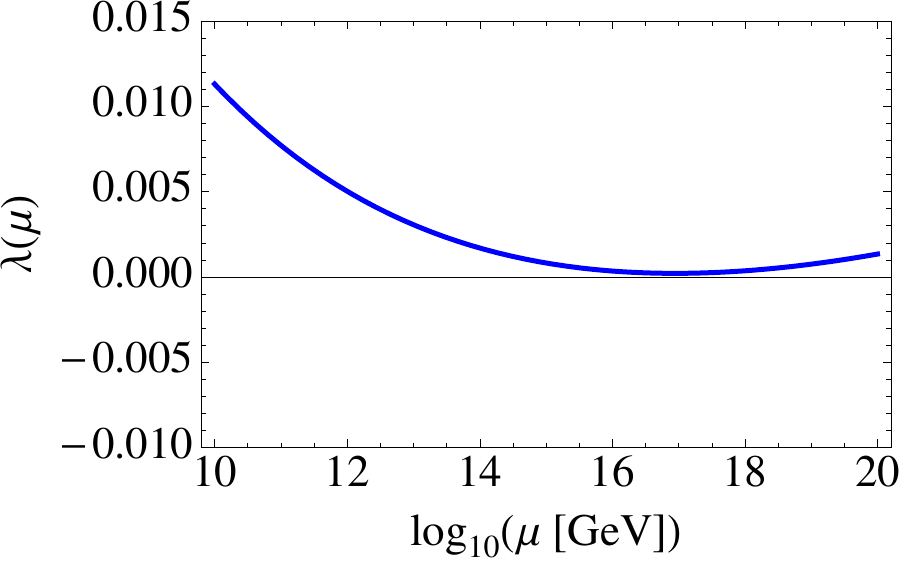}
\hfill\mbox{}
\caption{
Left: Three vertical blue lines represent the contour on which the $\lambda$ takes its minimum value at $10^{16}$, $10^{17}$, and $10^{18}$\GeV, respectively.
The lower red line gives the contour where the minimal value $\lambda_\text{min}$ becomes zero.
The upper green line  
gives the contour of $\lambda(10^{17}\GeV)=0$.
The width of the lines corresponds to varying $\rho_0$ from 0 to 0.6.
When we impose the stability up to e.g.\ $10^{17}$\GeV, the allowed region is below the lower red line in the right of the $\mu_\text{min}=10^{17}\GeV$ contour, and is below the upper green line in the left of the $\mu_\text{min}=10^{17}\GeV$ contour.
We see that varying $\rho_0$ does not change the stability bound much.
Right: RG flow of $\lambda(\mu)$ for $M_t=171.9\GeV$, $\kappa_0=0.14$, and $\rho_0=0.1$.
We see that $\lambda$ takes its minimum $\lambda_\text{min}=0$ around $10^{17}$GeV.
Other parameters are set to be $m_h=126\GeV$, $\alpha_S=0.1184$, and $m_W=80.384\GeV$.}\label{stability}
\end{center}
\end{figure}

The parameters $\kappa_0, \rho_0$ are not totally free but bounded by the perturbativity and stability of the theory:
\begin{itemize}
\item For perturbativity, it is required that $\kappa$ and $\rho$ cannot be too large up to the ultraviolet cutoff scale of the model. In this paper, we take this cutoff to be the typical string scale $\sim 10^{17}$\,GeV.
In Fig.~\ref{perturbativity}, we have shown the region where both $\kappa$ and $\rho$ are smaller than unity.
In this figure, the left of the two solid black lines corresponds to the weak coupling region.
This leads to the constraint:
\al{
\kappa_0
	&< 0.4, &
\rho_0
	&< 0.6.
}

\item
Next let us move on to the stability bound.
The parameter $\kappa$ does not change its sign through the RG running; $\rho$ monotonically increases; if $\rho_0\geq0$, this sector does not lead to an instability.
Therefore, the instability bound is simply that $\lambda$ keeps to be positive up to the cutoff scale $\Lambda$.
We have plotted this bound in the $M_t$ vs $\kappa_0$ plane in the left of Fig.~\ref{stability}.
We see that stability is recovered by introducing 
$\kappa_0$ of 0 to 0.3 depending on the top quark pole mass $M_t$.
In the right of Fig.~\ref{stability}, we plot $\lambda(\mu)$ for the case where $\lambda$ takes its minimum value 0 around $10^{17}\GeV$.
\end{itemize}

\begin{figure}[tn]
\begin{center}
\hfill
\includegraphics[width=.6\textwidth]{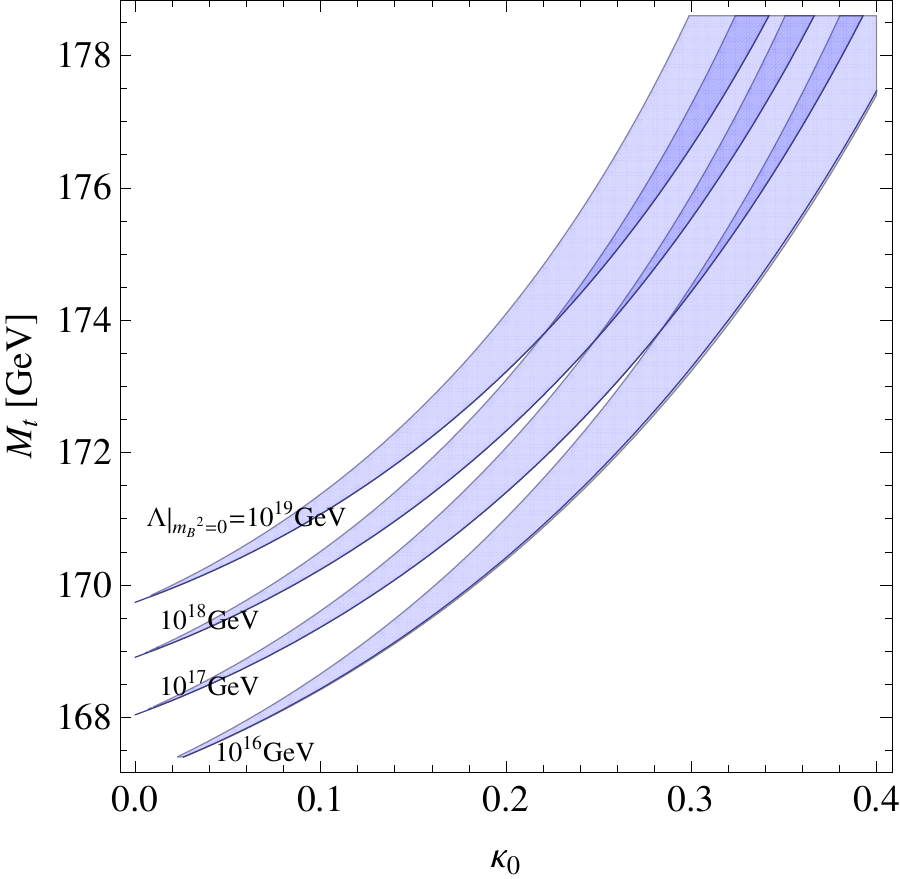}
\hfill\mbox{}
\caption{
Contours in $M_t$ vs $\kappa$ plane on which the bare Higgs mass becomes zero at $\Lambda=10^{16}$, $10^{17}$, $10^{18}$, and $10^{19}$\,GeV from below to above, respectively.
Width of each contour corresponds to varying $\rho_0$ from 0 to 0.6.
Other parameters are taken to be $m_h=126\GeV$, $\alpha_S=0.1184$, and $m_W=80.384\GeV$.}\label{Veltman}
\end{center}
\end{figure}

Let us examine how the fact that the bare Higgs mass becomes zero at the high scale is affected by inclusion of the extra $Z_2$ scalar.
At one-loop level, the bare Higgs mass becomes:
\al{
m_B^2&=-\left(6\lambda+\frac{3}{4}g_Y^2+\frac{9}{4}g_2^2-6y_t^2+\frac{1}{2}\kappa\right)I_1,
}
where the running couplings are evaluated at the cutoff scale $\Lambda$.
The $y_t^2$ term is the biggest for a low cutoff scale, whereas the gauge couplings and $\kappa$ dominate over $y_t^2$ for a larger cutoff scale; the sign is altered at an intermediate scale, where the bare mass becomes zero.
In Fig.~\ref{Veltman}, we plot the contour for $m_B^2=0$ in the $M_t$ vs $\kappa_0$ plane.

On the other hand, the bare mass for the newly added DM becomes
\al{
m_{S,B}^2&=-(2\kappa+\rho)I_1.
}
This is always negative and does not become zero for any cutoff scale.

\magenta{The bare masses become important when the temperature of the universe becomes close to $\Lambda$. The Hubble constant during the inflation gives the temperature of the de Sitter vacuum. At lower energies, the effective potential is given by the renormalized mass rather than the bare one, see e.g.\ Appendix B of Ref.~\cite{Hamada:2013mya}. Therefore these bare masses are irrelevant if $\Lambda$ is much larger than the Hubble scale.}

\section{Higgs inflation constraint}\label{constraint}
We discuss the constraint on the DM mass from the Higgs inflation above cutoff that we have proposed in Ref.~\cite{Hamada:2013mya}.

\subsection{Constraint from flat potential Higgs inflation}\label{MHI}
In this scenario, we focus on the fact that the SM Higgs potential becomes flat, namely the quartic coupling $\lambda$ and its beta function $\beta_\lambda$ becomes small, at high scales.
The RG improved Higgs potential is
\al{
V=\frac{1}{4}\lambda(h) h^4,
}
where we have chosen the renormalization scale $\mu=h$ for the running coupling in order to minimize the loop correction.

\begin{figure}[tn]
\begin{center}
\hfill
\includegraphics[width=.6\textwidth]{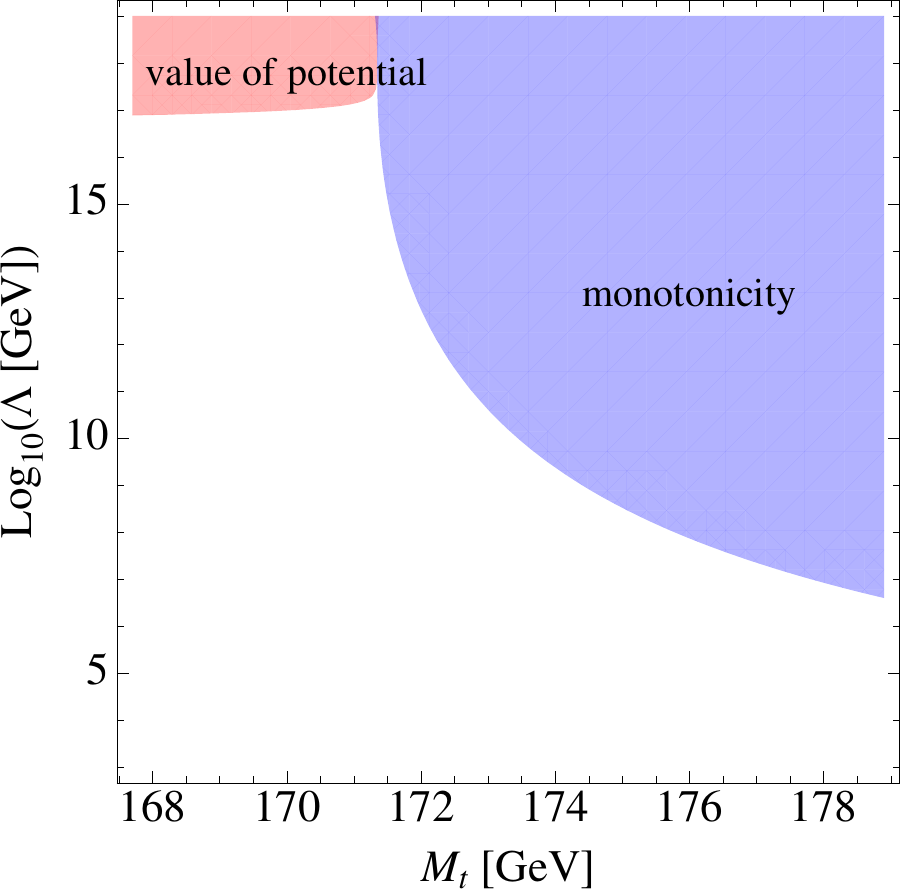}
\hfill\mbox{}
\caption{
The region excluded by the value of the potential (left, red) and by the monotonicity (right, blue) in the $\Lambda$ vs $M_t$ plane in order to realize the flat potential Higgs inflation within the SM.
}\label{SMbound}
\end{center}
\end{figure}

We can have a saddle point of this potential by tuning $M_t$, but this point cannot solely explain the small scalar perturbation; see e.g.\ Appendix~A of Ref.~\cite{Hamada:2013mya}.
In the Higgs inflation scenario we proposed earlier~\cite{Hamada:2013mya}, we assume that the Higgs potential becomes flat above the cutoff scale of the SM, leading to the inflation.
We do not know the shape of the potential above the cutoff scale, and there is a good chance that Nature realizes this scenario.
Since the prediction of the inflation depends on the physics beyond the cutoff scale, one may think that this is too arbitrary.
However, there are two constraints in order to achieve this scenario~\cite{Hamada:2013mya}:\footnote{
See Appendix~\ref{concrete_potential} for the possible potential above $\Lambda$.
}
\begin{itemize}
\item 
The detection of the scalar and tensor perturbation gives the upper bound on the value of the potential:
\al{\label{bound1}
V(\Lambda)<1.9\times 10^{65}\GeV^4\times\left(\frac{r}{0.16}\right),
}
where we have used the BICEP2 bound on the tensor-to-scalar ratio, $r=0.16^{+0.06}_{-0.05}$~\cite{Ade:2014xna}.
\item The monotonicity bound requires that the Higgs potential is monotonically increasing up to the cutoff $\Lambda$:
\al{\label{bound2}
{\df V\over\df h}
	&\geq	0
			\qquad\text{for $h\leq\Lambda$.}
}
\magenta{
This is required in order to avoid the so-called graceful exit problem in the end of the inflation~\cite{Hamada:2013mya}.
One may consider a possibility that there appears another positive minimum below $\Lambda$, and $\varphi$ goes over the barrier between this minimum and the electroweak one. As is stated in footnote~\ref{mono vs stab}, such a parameter region is narrow in the $M_t$-$\Lambda$ plane, and we neglect it; see e.g.\ Fig.~1 in Ref.~\cite{Hamada:2014iga}.
}
\end{itemize}
We plot these two bound in the case of the SM in Fig.~\ref{SMbound}.\footnote{\label{mono vs stab}
Parametrically the monotonicity and the stability ($\lambda>0$ for $h\leq\Lambda$) give almost identical results in the $\Lambda$ vs $M_t$ plane.
}

We note that in this scenario of the Higgs inflation above cutoff $\Lambda$, there is a lower bound on the tensor-to-scalar ratio $r$:
In the slow roll approximation, $r=16\epsilon_V$.
The magnitude of the scalar perturbation $A_s=V_*/24\pi^2M_P^4\epsilon_V$ is fixed to be $A_s=2.2\times10^{-9}$.
The potential height during the inflation $V_*$ should be larger than the potential height at the cutoff: $V_*>V(\Lambda)$.
In particular, we get
\al{
r> 10^{-3}
}
for $\Lambda\sim10^{17}\GeV$~\cite{Hamada:2013mya}.




\subsection{Dark matter mass prediction}

One of the most likely cutoffs is the string scale $\sim10^{17}$\,GeV.
The typical height of the potential is then
$V\sim\paren{10^{17}}^4\GeV^4=10^{68}\GeV^4$, which leads to too large tensor-to-scalar ratio to fit the observed cosmic microwave background spectrum.
However, $\lambda$ becomes small around $\Lambda$ for a range of $\kappa$, and we can evade this bounds.
To repeat, for the Higgs inflation above the cutoff, it is important that the discovered Higgs mass leads to very small $\lambda$ at high scales in the SM and in its extensions.

\begin{figure}[tn]
\begin{center}
\hfill
\includegraphics[width=.4\textwidth]{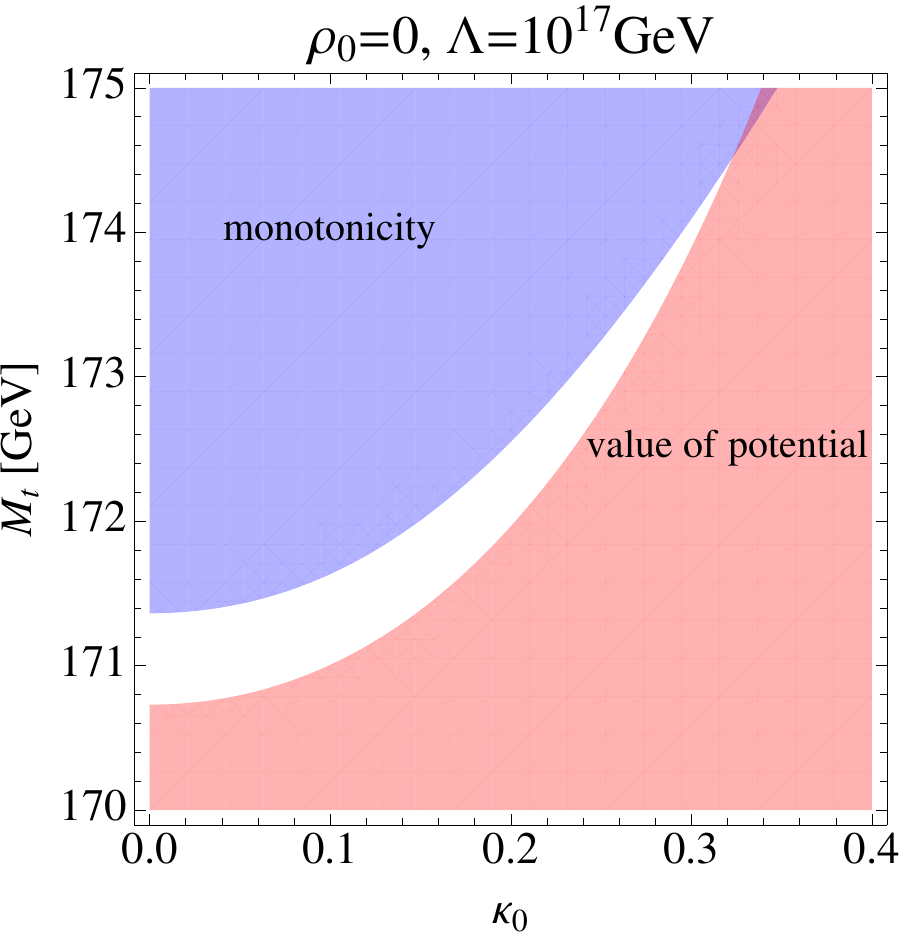}
\hfill
\includegraphics[width=.4\textwidth]{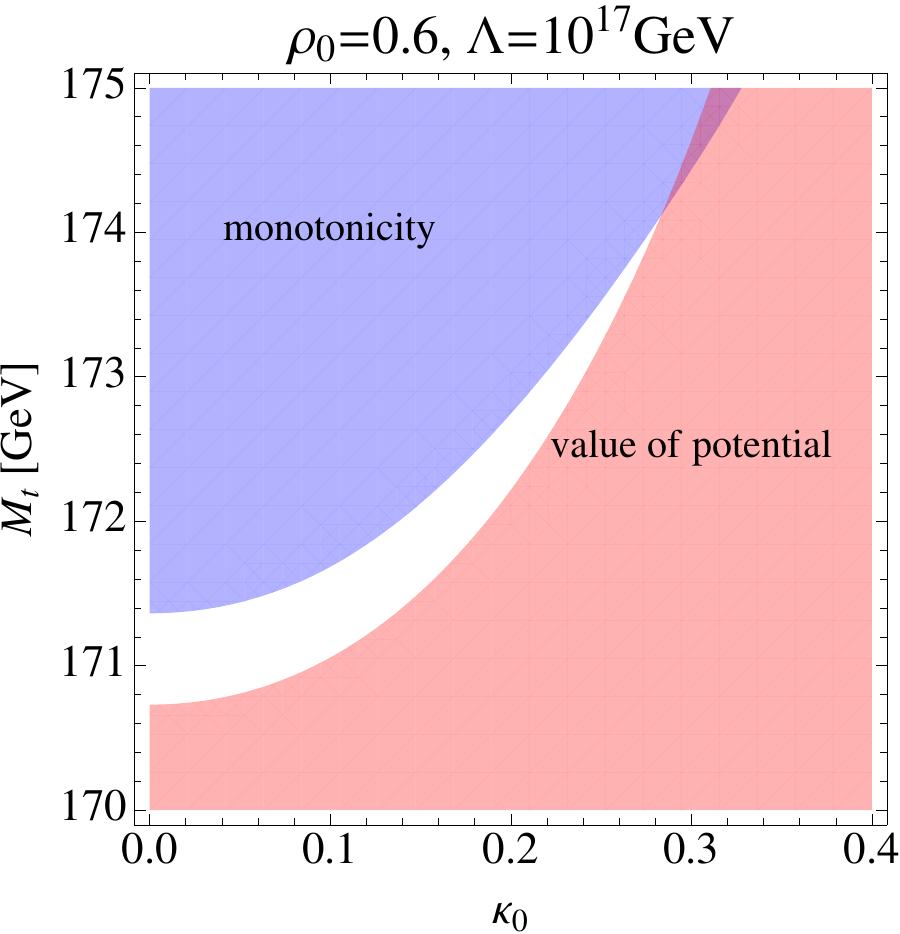}
\hfill\mbox{}
\caption{
The excluded regions in the $M_t$ vs $\kappa$ plane from the monotonicity (upper-left, blue) and from the value of the potential (lower-right, red) in the $Z_2$ scalar DM model to achieve the flat potential Higgs inflation.
Left and right panels are for $\rho=0$ and 0.6, respectively.
}
\label{Higgs_inflation_kappa}
\end{center}
\end{figure}

\begin{figure}[tn]
\begin{center}
\hfill
\includegraphics[width=.4\textwidth]{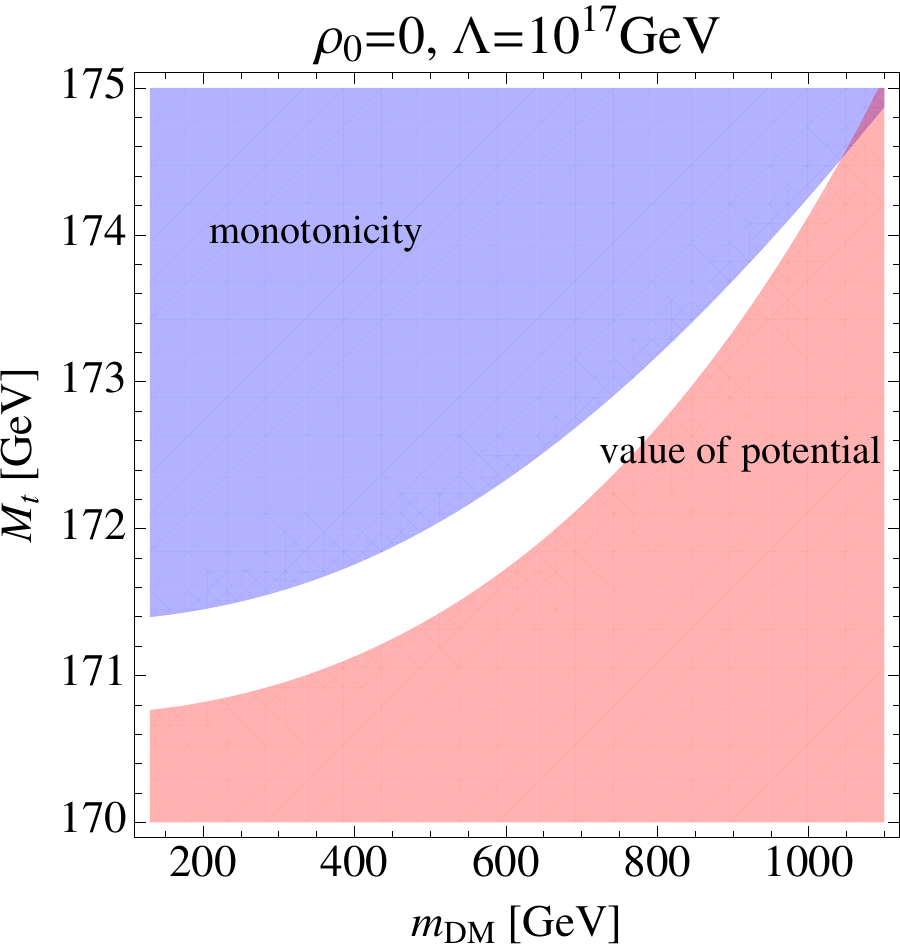}
\hfill
\includegraphics[width=.4\textwidth]{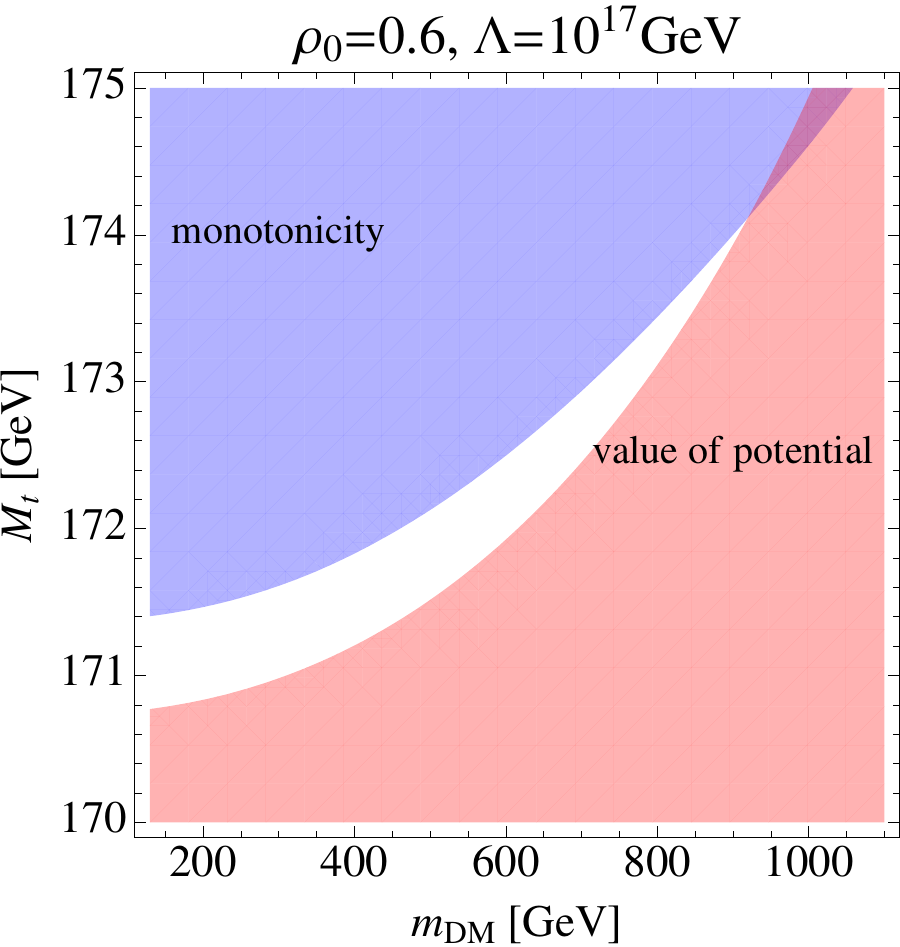}
\hfill\mbox{}
\caption{
The excluded regions in the $M_t$ vs $m_\text{DM}$ plane from the monotonicity (upper-left, blue) and from the value of the potential (lower-right, red) in the $Z_2$ scalar DM model to achieve the flat potential Higgs inflation.
Left and right panels are for $\rho_0=0$ and 0.6, respectively.
}
\label{Higgs_inflation_mDM}
\end{center}
\end{figure}

Let us show that the DM mass is strongly constrained from the conditions explained in Sec.~\ref{MHI}.
In the left (right) of Fig.~\ref{Higgs_inflation_kappa}, we plot the excluded regions in the $M_t$ vs $\kappa$ plane for the cutoff scale $\Lambda=10^{17}$\,GeV and for $\rho_0=0$ (0.6). 
We see that the resultant constraint is $\kappa_0< 0.32$ and $171\GeV< M_t< 174\GeV$. 
By using Eq.~\eqref{thermal},
this allowed region for $\kappa$ translates to the allowed region for the DM mass shown in Fig.~\ref{Higgs_inflation_mDM}:
\al{\label{DMmass1}
m_\text{DM}<1000\GeV.
}
In Fig.~\ref{Higgs_inflation_mDM}, we can see that there is a strong correlation between the top and DM masses. The relation between $\kappa$ and $m_\text{DM}$ is rather complicated for $\kappa<0.04 \ (m_\text{DM}< 130\GeV)$~\cite{Cline:2013gha}, and we show the region where $m_\text{DM}> 130\GeV$ in Fig.~\ref{Higgs_inflation_mDM}.

\begin{figure}[tn]
\begin{center}
\hfill
\includegraphics[width=.32\textwidth]{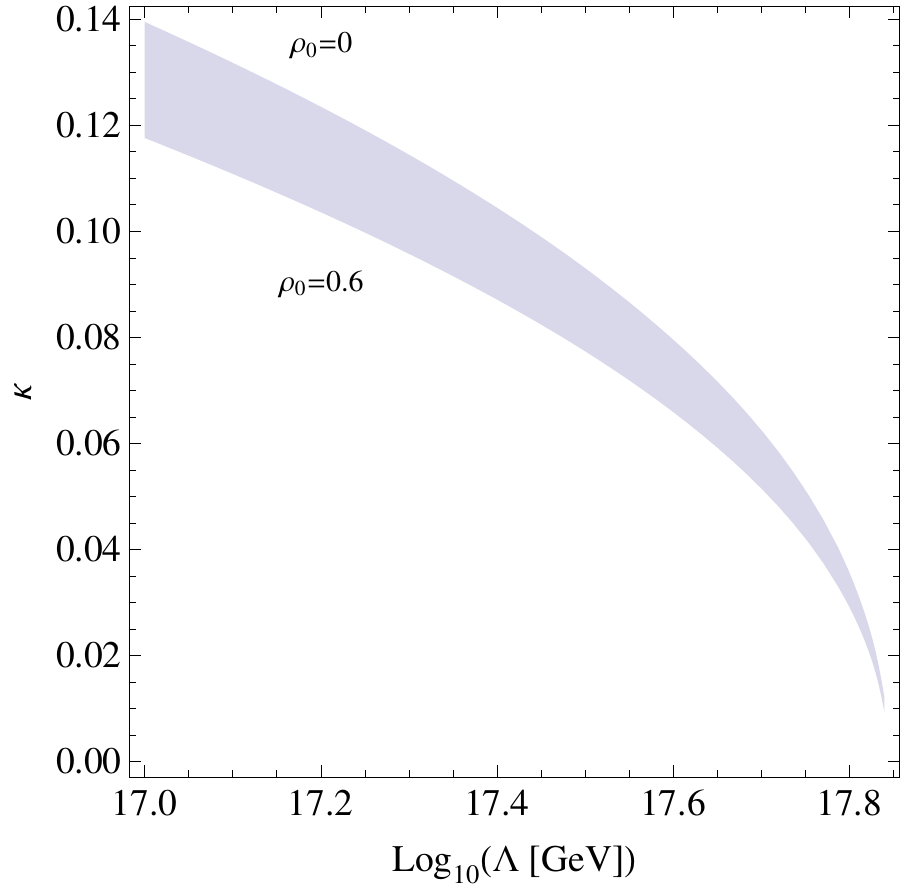}
\hfill
\includegraphics[width=.32\textwidth]{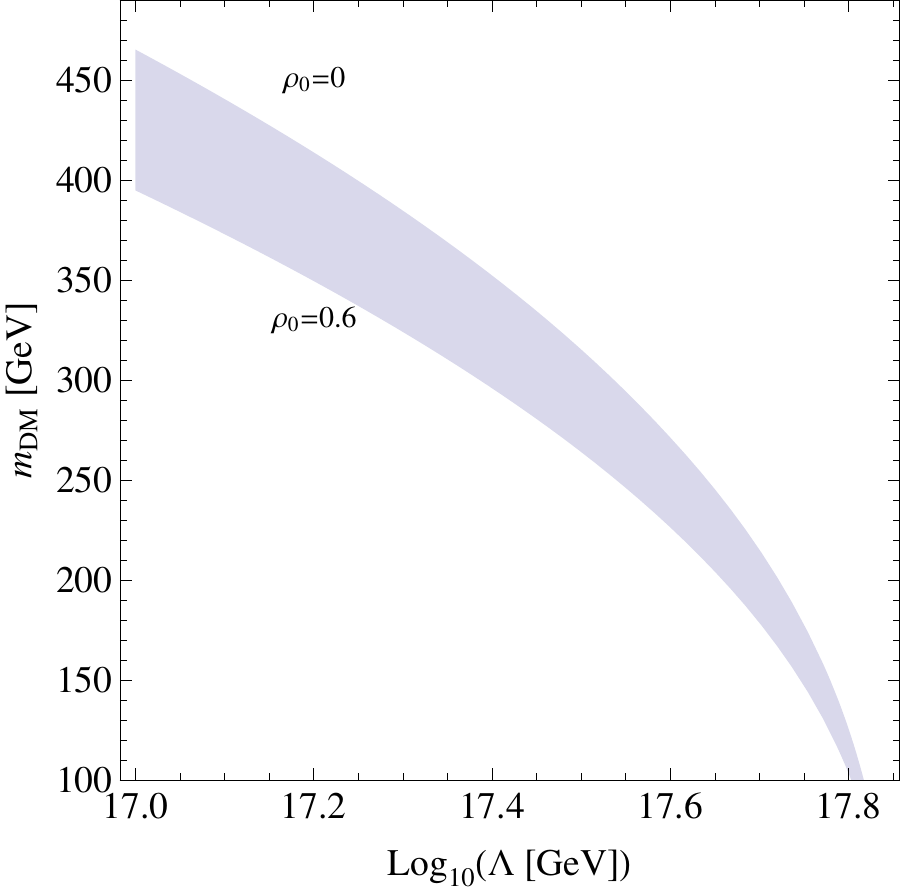}
\hfill
\includegraphics[width=.32\textwidth]{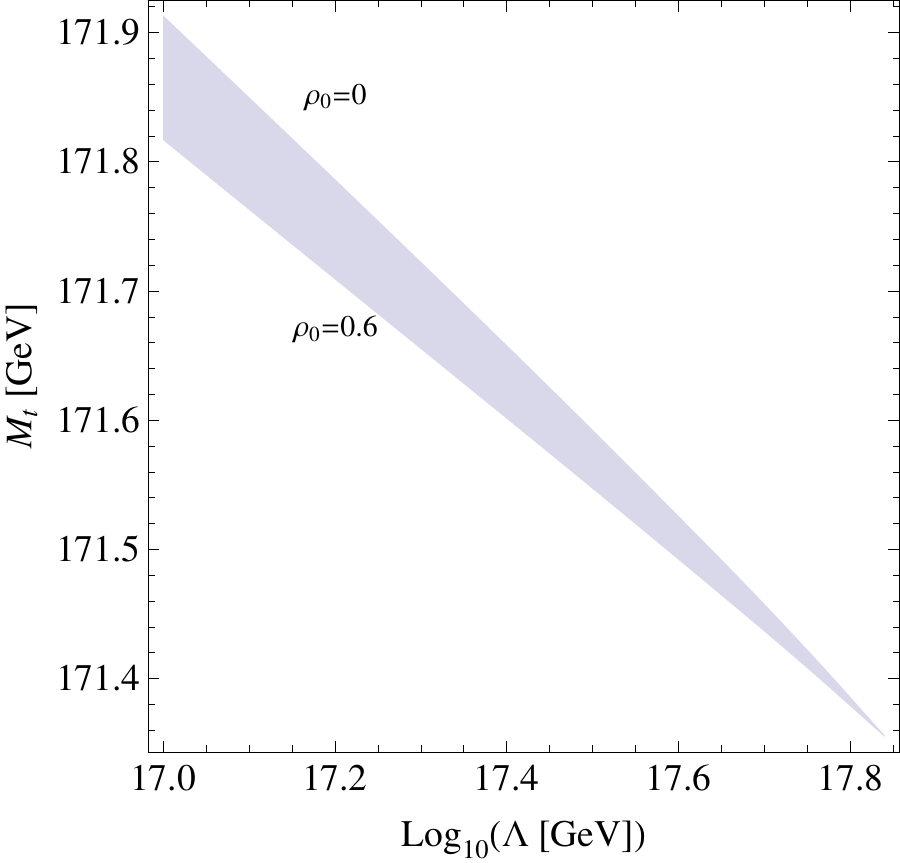}
\hfill\mbox{}
\caption{
$\kappa$ (left), $m_\text{DM}$ (center), and $M_t$ (right) as functions of $\Lambda$ under the flatness assumption $\lambda(\Lambda)\sim\beta_\lambda(\Lambda)\sim0$. The band shows the range $0\leq\rho_0\leq0.6$.
Higgs mass is taken to be $m_H=126\GeV$.
}
\label{mDM_vs_Lambda}
\end{center}
\end{figure}

\begin{figure}[tn]
\begin{center}
\hfill
\includegraphics[width=.4\textwidth]{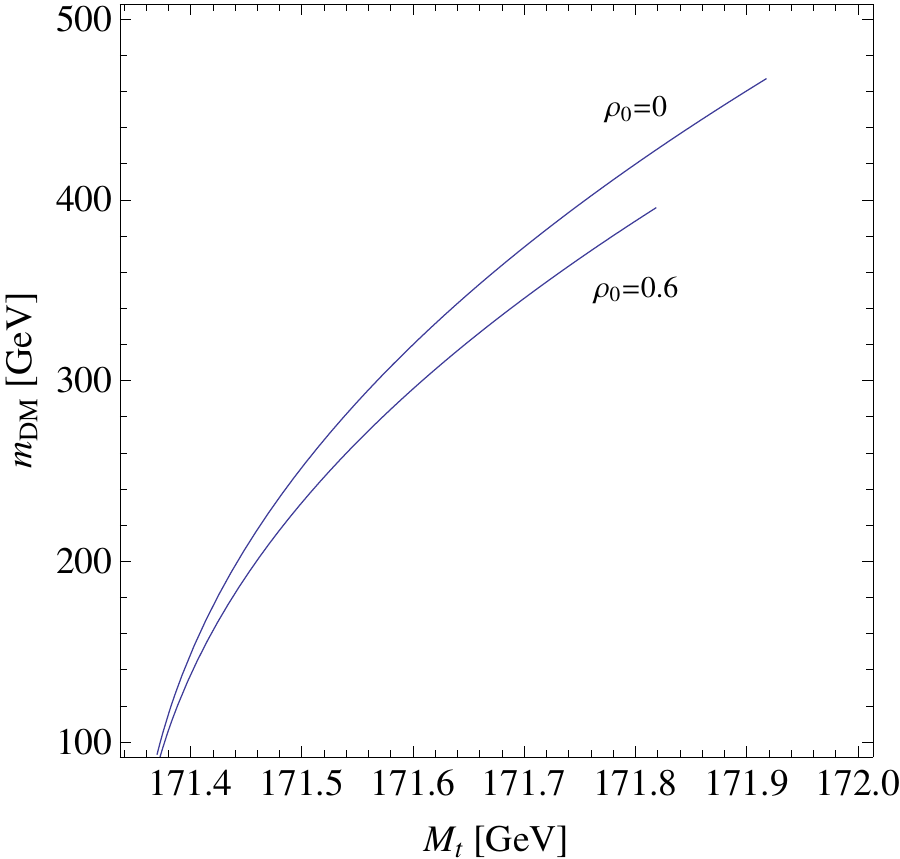}
\hfill\mbox{}
\caption{
$m_\text{DM}$ vs $M_t$ with the same condition as in Fig.~\ref{mDM_vs_Lambda}.
The range $\Lambda>10^{17}\GeV$ is plotted.
}
\label{mDM_vs_Mt}
\end{center}
\end{figure}

The MPP~\cite{Froggatt:1995rt,Froggatt:2001pa,Nielsen:2012pu} and the Higgs inflation at the critical point~\cite{Hamada:2014iga,Bezrukov:2014bra} both requires the flatness of the SM potential: $\lambda\sim\beta_\lambda\sim0$ around $\Lambda\sim10^{17}\GeV$.\footnote{
With the MPP, we impose $\lambda(\Lambda)=\beta_\lambda(\Lambda)=0$, while for the Higgs inflation at the critical point, we need $\displaystyle\lambda(\Lambda)={1\over32}{\df\beta_\lambda\over\df\ln\mu}(\Lambda)$ at the $\Lambda$ where $\beta_\lambda(\Lambda)=0$; see e.g.\ Ref.~\cite{Hamada:2014iga}. Parametrically both conditions give the same result.
}
When we impose this flatness condition, the coupling $\kappa$, the DM mass $m_\text{DM}$, and the top mass $M_t$ are completely fixed as functions of $\rho_0$:
\al{
0.12	&\leq	\kappa_0	\leq	0.14,	\\
400\GeV	&\leq	m_\text{DM}\leq 470\GeV,	\label{DM mass}\\
171.8\GeV
		&\leq	M_t	\leq 171.9\GeV,
}
where the lower and upper ends correspond to the parameter choice $\rho_0=0$ and 0.6, respectively. Here we have chosen $m_h=126\GeV$; the result is insensitive to the Higgs mass compared to the top mass.
In Fig.~\ref{mDM_vs_Lambda}, we plot $\kappa_0$, $m_\text{DM}$, and $M_t$ as functions of $\Lambda$; again the lower and upper ends correspond to the parameter choice $\rho_0=0$ and 0.6, respectively.

\magenta{
To be specific, we quote the numbers for the spin independent cross section for the DM-nucleon scattering~\cite{Cline:2013gha}:
\al{
\sigma_\text{SI}
	&=	9.5\times10^{-46}\,\text{cm}^2
			\sqbr{\lambda\over0.05}^2
			\sqbr{200\,\text{GeV}\over m_\text{DM}}^2
	=	8.4\times10^{-46}\,\text{cm}^2,
		\label{SI cross section}
}
where we have put Eq.~\eqref{thermal} in the last step.
}

Both of the DM mass predictions~\eqref{DMmass1} and \eqref{DM mass} \magenta{with the cross section~\eqref{SI cross section}} are well within the reach of the future DM detection experiments such as XENON1T~\cite{Aprile:2012zx} and LUX~\cite{Akerib:2013tjd}.
This scenario can also be tested by the strong correlation between the DM and top masses as shown in Fig.~\ref{Higgs_inflation_mDM} and Fig.~\ref{mDM_vs_Mt}.

From Fig.~\ref{mDM_vs_Lambda}, we see that we cannot get a flat potential for $\Lambda>10^{17.8}\GeV$.
This is because the larger the $\Lambda$ is, the smaller the $\kappa_0$ becomes.
$\kappa_0\to0$ is the SM limit.
In the SM, the scale of the flat potential, obtained by tuning $M_t=171.1\GeV$, is $10^{17.8}\GeV$.
When the DM decouples from the Higgs, it becomes impossible to raise the Higgs potential against the top Yukawa contribution.

If one wants, one can get the flat potential at higher $\Lambda>10^{17.8}\GeV$ by introducing the right handed neutrino.
For simplicity, we consider the case where one neutrino Yukawa coupling is large.
This coupling contributes to the RG running of $\lambda$ above the Majarana mass similarly to the top Yukawa coupling.
Therefore, balancing the contribution from the scalar DM and the right-handed neutrino, we can obtain the flat potential even above $10^{17.8}$\,GeV.
For example, we get $m_\text{DM}=820\GeV$ for $M_t=173\GeV$, $\Lambda=10^{18}\GeV$, $m_h=126\GeV$, and $\rho_0=0.1$.
See Appendix~\ref{neutrino} for details.

\section{Conclusions}\label{summary}
We have obtained the constraints on the DM mass $m_\text{DM}<1000\GeV$ and on the top quark mass $171\GeV< M_t< 174\GeV$ in the Higgs portal $Z_2$ singlet model, imposing the condition to achieve the Higgs inflation above the cutoff~\cite{Hamada:2013mya}.
We can further fix the DM mass to be $400\GeV< m_\text{DM}<470\GeV$ if we impose the flatness of the potential around the string scale $10^{17}\GeV$, $\lambda\sim\beta_\lambda\sim0$, as is required by the multiple point principle or by the Higgs inflation at the critical point.
Both of these regions are testable in the near future by the direct and indirect DM detection experiments and by the precision measurement of the top quark pole mass.

We emphasize that the idea of the Higgs inflation above cutoff uses the fact the Higgs quartic coupling and its beta function become very close to zero around the string scale $10^{17}\GeV$.
Therefore, if the relation between the DM and top masses is confirmed, it suggests that the cutoff of the SM is around $10^{17}\GeV$.\footnote{
See e.g.\ Ref.~\cite{Mizoguchi:2014gva,Hamada:2014hpa,Hamada:2014eia} for an attempt along this line.
}

This framework does not predict a particular value of the spectral index $n_s$, but gives the tensor-to-scalar ratio $r> 10^{-3}$.
This prediction is consistent to the B-mode polarization detection by the BICEP2 experiment. 
It would be interesting to construct a model that makes the Higgs potential above the cutoff to be flat in field theory and in string theory~\cite{Hamada:2015ria}.

\subsection*{Note added}
While this paper was in preparation, there appeared Ref.~\cite{Haba:2014zda} with similar subject. The result is in qualitative agreement with ours.
Afterwards, there also appeared a related work~\cite{Channuie:2014kda}.

\subsection*{Acknowledgement}
We thank Seong Chan Park for useful discussions and Yukinari Sumino for helpful comment.
This work is in part supported by the Grant-in-Aid for Scientific Research Nos.~22540277 (HK), 23104009, 20244028, and 23740192 (KO). The work of Y. H. is supported by a Grant-in-Aid for Japan Society for the Promotion of Science (JSPS) Fellows No.25$\cdot$1107. 
\appendix
\section{Renormalization group equations}\label{RGE}
We summarize the two-loop RGEs:\footnote{
We calculate two-loop RGEs by using the results of Refs.~\cite{Machacek:1983tz,Machacek:1983fi,Machacek:1984zw}.
After completion, there appeared Ref.~\cite{HabaNew} which also computes the two-loop RGEs.
By putting $N=1$ and $y_\nu=0$ in our and their results, respectively, RGEs become the same.
}
\begin{align}
\frac{dg_Y}{dt}
	&=	{1\over16\pi^2}\frac{41}{6}g_Y^3+\frac{g_Y^3}{(16\pi^2)^2}\left({199\over18}g_Y^2+{9\over2}g_2^2+{44\over3}g_3^2-{17\over6}y_t^2\right),
	\nn
\frac{dg_2}{dt}&=-{1\over16\pi^2}\frac{19}{6}g_2^3
	+\frac{g_2^3}{(16\pi^2)^2}\left({3\over2}g_Y^2+{35\over6}g_2^2+12g_3^2-{3\over2}y_t^2\right),
	\nn
\frac{dg_3}{dt}
	&=	-\frac{7}{16\pi^2}g_3^3+\frac{g_3^3}{(16\pi^2)^2}\left({11\over6}g_Y^2+{9\over2}g_2^2-26g_3^2-2y_t^2\right),
	\nn
\frac{dy_t}{dt}
	&=	\frac{y_t}{16\pi^2}\bigg(\frac{9}{2}y_t^2-\frac{17}{12}g_Y^2-\frac{9}{4}g_2^2-8g_3^2\bigg)+\frac{y_t}{(16\pi^2)^2}\bigg(-12y_t^4+6\lambda^2
	+
	\frac{1}{4}N\kappa^2
	-12\lambda y_t^2\nn
	&\quad
		+\frac{131}{16}g_Y^2 y_t^2+\frac{225}{16}g_2^2 y_t^2+36 g_3^2 y_t^2+\frac{1187}{216}g_Y^4-\frac{23}{4}g_2^4-108g_3^4-\frac{3}{4}g_Y^2 g_2^2+9g_2^2 g_3^2+\frac{19}{9}g_3^2 g_Y^2\bigg),
		\nn
\frac{d\lambda}{dt}
	&=	\frac{1}{16\pi^2}\bigg(
	\frac{1}{2}N\kappa^2
	+24 \lambda^2-3g_Y^2 \lambda-9g_2^2 \lambda+\frac{3}{8}g_Y^4+\frac{3}{4}g_Y^2 g_2^2 +\frac{9}{8}g_2^4+12\lambda y_t^2-6y_t^4\bigg)\nn
	&\quad
		+\frac{1}{(16\pi^2)^2}\bigg\{
-2N\kappa^3-5N\kappa^2\lambda
			-312\lambda^3+36\lambda^2(g_Y^2+3g_2^2)
			-\lambda\left(-{629\over24}g_Y^4-{39\over4}g_Y^2g_2^2+{73\over8}g_2^4\right)\nn
	&\phantom{\quad+\frac{1}{(16\pi^2)^2}\bigg\{}
		+\frac{305}{16}g_2^6-\frac{289}{48}g_Y^2 g_2^4 -\frac{559}{48}g_Y^4 g_2^2 -\frac{379}{48}g_Y^6 -32 g_3^2 y_t^4-\frac{8}{3}g_Y^2 y_t^4-\frac{9}{4}g_2^4 y_t^2\nn
	&\phantom{\quad+\frac{1}{(16\pi^2)^2}\bigg\{}
		+\lambda y_t^2 \bigg(\frac{85}{6}g_Y^2+\frac{45}{2}g_2^2+80g_3^2\bigg)+g_Y^2 y_t^2\bigg(-\frac{19}{4}g_Y^2+\frac{21}{2}g_2^2\bigg)\nn
	&\phantom{\quad+\frac{1}{(16\pi^2)^2}\bigg\{}
		-144 \lambda^2 y_t^2-3\lambda y_t^4+30y_t^6\bigg\},
		\nn
\frac{d\kappa}{dt}	
        &=
          \frac{\kappa}{16\pi^2}\left(
          12\lambda+ \rho+\frac{N-1}{3}\rho+4\kappa+6y_t^2-\frac{3}{2}g_Y^2 -\frac{9}{2}g_2^2  
          \right) \nn
&
+\frac{\kappa}{(16\pi^2)^2}\bigg\{
-\left(\frac{N}{2}+10\right)\kappa^2-72\kappa\lambda-60\lambda^2-(2N+4)\kappa\rho-\left(\frac{5N+10}{18}\right)\rho^2\nn
&
-y_t^2(12\kappa +72\lambda)
-\frac{27}{2} y_t^4
+g_Y^2 (\kappa +24\lambda )
+ g_2^2(3\kappa+72\lambda)
+y_t^2\left(\frac{85}{12}  g_Y^2+\frac{45}{4} g_2^2+40 g_3^2\right)
\nn
&
+\frac{557}{48}g_Y^4
-\frac{145}{16}g_2^4
+\frac{15}{8}g_Y^2 g_2^2
\bigg\},
\nn
\frac{d\rho}{dt}	
        &=\frac{1}{16\pi^2}\left(3\rho^2
        +
        \frac{N-1}{3}\rho^2
        +12 \kappa^2\right)
+\frac{1}{(16\pi^2)^2}\bigg\{
-\frac{3N+14}{3}\rho^3-20\kappa^2 \rho-48\kappa^3\nn
&
-72\kappa^2 y_t^2
+24\kappa^2 g_Y^2
+72\kappa^2 g_2^2
\bigg\},
\end{align}
where $t=\ln\mu$ and $N$ is the number of the real singlet scalar. 
The model we considered in the paper is $N=1$.
The general $N$ case corresponds to the following replacement in the Lagrangian \eqref{Lagrangian},
\al{
S^2\rightarrow \sum_{i=1}^N S_i^2.
}
When we include right handed neutrino, the one-loop RGEs are given by
\begin{align}
\frac{d\lambda}{dt}
	&=	\frac{1}{16\pi^2}\bigg(\frac{1}{2}\kappa^2+24 \lambda^2-3g_Y^2 \lambda-9g_2^2 \lambda+\frac{3}{8}g_Y^4+\frac{3}{4}g_Y^2 g_2^2 +\frac{9}{8}g_2^4+12\lambda y_t^2+4\lambda y_{\nu}^2-6y_t^4-2y_{\nu}^4\bigg),\nn
\frac{d\kappa}{dt}	
        &=\frac{1}{16\pi^2}\left(12\kappa\lambda+\kappa \rho+4\kappa^2+6y_t^2\kappa-\frac{3}{2}g_Y^2 \kappa-\frac{9}{2}g_2^2 \kappa+2y_{\nu}^2 \kappa \right),\nn
\frac{d\rho}{dt}	
        &=\frac{1}{16\pi^2}\left(3\rho^2+12 \kappa^2\right),\nn
        \frac{dy_{\nu}}{dt}
        &=\frac{y_{\nu}}{16\pi^2}\left(\frac{5}{2}y_{\nu}^2+3y_t^2-\frac{9}{4}g_2^2-\frac{3}{4} g_Y^2\right).
        	\label{RGEs}
\end{align}
The one-loop bare mass is 
\al{
m_B^2&=-\paren{6\lambda+\frac{3}{4}g_Y^2+\frac{9}{4}g_2^2-6y_t^2+\frac{1}{2}\kappa-2y_\nu^2}I_1,\\
m_{S B}^2&=-\paren{2\kappa+\rho}I_1.
}

\section{Concrete potential above cutoff}
\label{concrete_potential}
Here we show some examples of the concrete potential above cutoff.
\begin{itemize}
\item
In Ref.~\cite{Hamada:2013mya}, we have shown the result for the log potential as a concrete example.
Indeed, the Coleman-Weinberg potential in the cutoff theory leads to a log potential in the region where field value exceeds the cutoff; see Appendix B of Ref.~\cite{Hamada:2013mya}.
That is, above $\Lambda$ we get
\al{
V=V_0+V_1\ln\frac{h}{\Lambda}.
}
This potential predicts $n_s\simeq0.98$ and $r\simeq10^{-2}$, which requires further modification because of the recent tensor mode result from the BICEP2 experiment~\cite{Ade:2014xna}. 

\item
In the spirit of Ref.~\cite{Hamada:2013mya}, we can make use of the flat Higgs potential, and force it to be further flat above $\Lambda\sim M_P/\xi$ by introducing the non-minimal coupling $\xi$ between the Higgs and Ricci scalar.
While the original version of the Higgs inflation~\cite{Bezrukov:2007ep} predicts too small value of $r$ to be consistent with the BICEP2 result, the above-stated strategy leads to sufficiently large $r$~\cite{Hamada:2014iga,Bezrukov:2014bra}; see also \cite{Allison:2013uaa,Cook:2014dga,Hamada:2014wna,Hamada:2014raa}.

\item If we get above the cutoff $\Lambda$
\al{
V=c \Lambda^2 h^2,
}
with $c$ being a coefficient, then we obtain the chaotic inflation with the quadratic potential~\cite{Linde:1983gd}.
This is known to be consistent to the current observation of $n_s$ and $r$~\cite{Ade:2013uln,Ade:2014xna}.
This kind of quadratic potential is realized if the bare mass
takes a tiny non-zero value after integrating out the heavy modes, while all other terms are much suppressed. More precisely, 
the Coleman-Weinberg potential in the bare perturbation theory is given by
\al{
V(h)
	&=	{m_B^2\over 2}h^2+{\lambda_B\over 4}h^4
		+\sum_i{N_i\over2}\int{\df^4p\over(2\pi)^4}\ln{p^2+c_ih^2\over p^2},
		\label{bare_potential}
}
where $N_i$ is the number of degrees of freedom, $c_i$ is the coupling to the Higgs, and $\lambda_B$ is the bare Higgs self coupling. If the top mass is 170 to 171\GeV, $m_B^2$ and $\lambda_B$ become almost zero for the cutoff scale around $10^{17\text{--}18}$\GeV~\cite{Hamada:2012bp}.
Further it is possible that $m_B^2$ is very small but non-zero and that $\lambda_B=0$.
Then Higgs quadratic term dominates the bare potential~\eqref{bare_potential}. 
The asymptotic form of the potential becomes 
\al{
V(h)={m_B^2\over 2}h^2,
}
and we obtain quadratic chaotic inflation.
\end{itemize}

\section{Right-handed neutrino}\label{neutrino}

\begin{figure}[tn]
\begin{center}
\hfill
\includegraphics[width=.4\textwidth]{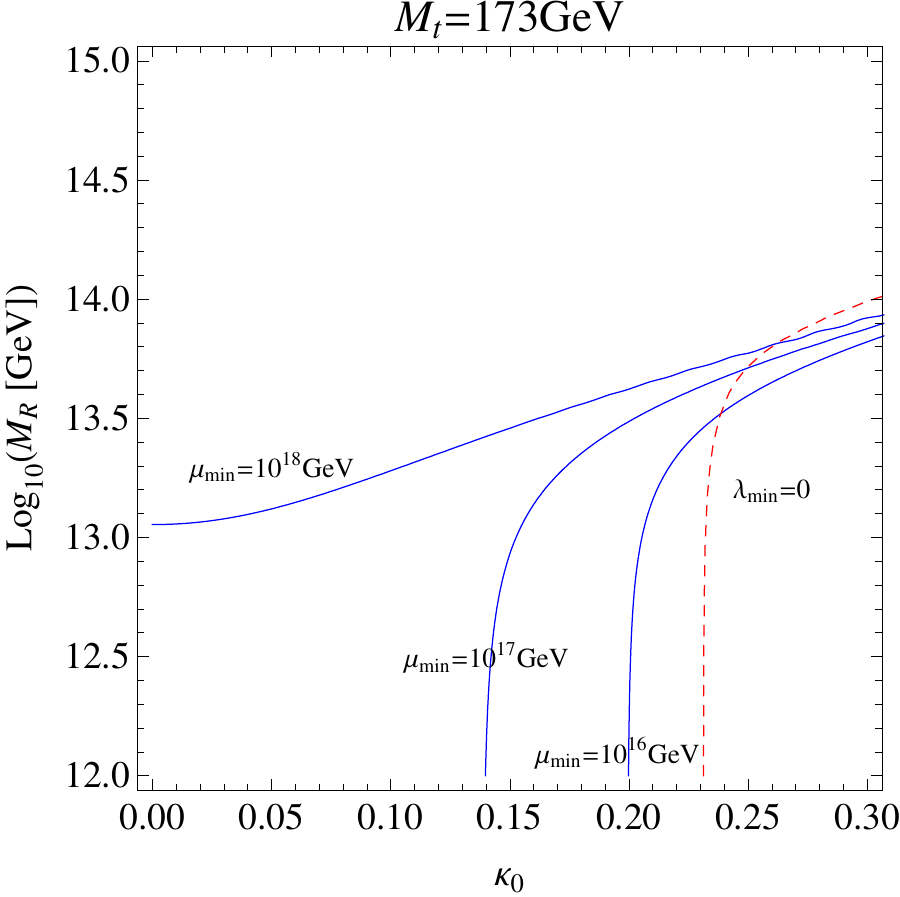}
\hfill\mbox{}
\caption{With the right-handed neutrino, we plot in $M_R$ vs $\kappa_0$ plane the contours with $\mu_\text{min}=10^{16}\GeV$, $10^{17}\GeV$, and $10^{18}\GeV$ by blue lines from below to above, respectively, where $\mu_\text{min}$ is the scale at which $\lambda(\mu)$ takes its minimum value, $\beta_\lambda(\mu_\text{min})=0$.
We also plot a contour of $\lambda_\text{min}=0$ by the dashed (red) line, where $\lambda_\text{min}=\lambda(\mu_\text{min})$.
Here $y_\nu$ is chosen such that the seesaw mass of the neutrino becomes 0.1\,eV.
Other parameters are fixed to be $m_h=126\GeV$, $M_t=173\GeV$, and $\rho_0=0.1$.
}\label{neutrino_fig}
\end{center}
\end{figure}

Let us examine the constraint on the Higgs portal $Z_2$ scalar model when we add the right handed neutrinos, in order to achieve the flat potential above $10^{17.8}\GeV$ and/or with a heavier top mass $M_t>172\GeV$. 
For simplicity, we add only one generation of neutrino:
\al{
\Delta \mathcal{L}=\bar{\nu}_R i \gamma^\mu\partial_\mu \nu_R-M_R \bar{\nu}_R^c \nu_R
-\paren{y_\nu\bar{L}H^\dagger \nu_R+\text{h.c.}}.
}
Then the beta functions for $y_t$, $\lambda$ and $\kappa$ are modified at the one-loop level:
\al{
\frac{dy_t}{dt}&=\frac{y_t}{16\pi^2}
\left(\frac{9}{2}y_t^2+y_\nu^2-\frac{17}{12}g_Y^2-\frac{9}{4}g_2^2-8g_3^2\right),\nn
\frac{d\lambda}{dt}
	&=	\frac{1}{16\pi^2}\bigg(\frac{1}{2}\kappa^2+24 \lambda^2-3g_Y^2 \lambda-9g_2^2 \lambda+\frac{3}{8}g_Y^4+\frac{3}{4}g_Y^2 g_2^2 +\frac{9}{8}g_2^4+12\lambda y_t^2+4\lambda y_{\nu}^2-6y_t^4-2y_{\nu}^4\bigg),\nn
\frac{d\kappa}{dt}	
        &=\frac{1}{16\pi^2}\left(12\kappa\lambda+\kappa \rho+4\kappa^2+6y_t^2\kappa-\frac{3}{2}g_Y^2 \kappa-\frac{9}{2}g_2^2 \kappa+2y_{\nu}^2 \kappa \right).
}
The beta function for $y_\nu$ reads
\al{
\frac{dy_{\nu}}{dt}
        &=\frac{y_{\nu}}{16\pi^2}\left(\frac{5}{2}y_{\nu}^2+3y_t^2-\frac{9}{4}g_2^2-\frac{3}{4} g_Y^2\right).
}
The modification due to the inclusion of the neutrino is in effect above the right handed Majorana mass scale, $\mu=M_R$.
We see that $y_\nu$ contribute negatively to the RG running of $\lambda$, similarly to $y_t$. 
This effect is opposite to the inclusion of the scalar DM.
Balancing these two, we can get the flat potential above $10^{17.8}\GeV$ and/or with a heavier top mass $M_t>172\GeV$, as is plotted in Fig.~\ref{neutrino_fig}.
Here we have chosen the Yukawa $y_\nu$ such that the neutrino mass becomes 0.1\,eV. That is, we imposed
\al{
\frac{y_\nu^2 v^2}{2 M_R}=0.1\eV,
\quad
\text{with}
\quad
v\simeq246\GeV,
}
at the scale $\mu=M_R$.
We see from Fig.~\ref{neutrino_fig} that rather large $\kappa\sim 0.25$ can realize the flat potential.
This value of $\kappa$ corresponds to the dark matter mass 820\,GeV.


\bibliographystyle{TitleAndArxiv}
\bibliography{HKO}

\end{document}